\documentclass[showpacs,pra,a4paper,floatfix,twocolumn,superscriptaddress]{revtex4}
\usepackage{graphicx}
\usepackage{bm,epsfig}
\usepackage{dsfont}
\usepackage{amsmath}
\usepackage{amssymb}

\newcommand{\ket}[1]{\ensuremath{| #1\rangle}}
\newcommand{\bra}[1]{\ensuremath{\langle #1 |}}
\newcommand{\com}[2]{\ensuremath{\left[ #1, #2 \right]}}

\allowdisplaybreaks[2]

\begin{document}

\title{Phase modulation induced by cooperative effects \\
in electromagnetically induced transparency}

\author{Robert \surname{Fleischhaker}}
\affiliation{Max-Planck-Institut f\"ur Kernphysik, Saupfercheckweg 1, 
D-69117 Heidelberg, Germany}

\author{Tarak N. \surname{Dey}}
\affiliation{Max-Planck-Institut f\"ur Kernphysik, Saupfercheckweg 1, 
D-69117 Heidelberg, Germany}
\affiliation{Indian Institute of Technology Guwahati, 
Guwahati- 781 039, Assam, India}

\author{J\"org \surname{Evers}}
\affiliation{Max-Planck-Institut f\"ur Kernphysik, Saupfercheckweg 1, 
D-69117 Heidelberg, Germany}

\date{\today}

\begin{abstract}
We analyze the influence of dipole-dipole interactions in an electromagnetically induced transparency setup for a density at the onset of cooperative effects. To this end, we include mean-field models for the influence of local field corrections and radiation trapping into our calculation. We show both analytically and numerically that the polarization contribution to the local field strongly modulates the phase of a weak pulse. We give an intuitive explanation for this {\em local field induced phase modulation} and demonstrate that it distinctively differs from the nonlinear self-phase modulation a strong pulse experiences in a Kerr medium.
\end{abstract}


%
%

\maketitle

\section{\label{secint}Introduction}
Electromagnetically induced transparency (EIT) stands out as one of the most useful coherence and interference phenomena (see \cite{eit1,eit2} and references therein).
Current research focuses on dilute samples with $N \lambda^3 \ll 1$ ($N$ density, $\lambda$ transition wavelength), in which the atoms essentially act independently.
%
Experimentally, an important reason for the restriction to low densities is detrimental decoherence induced, e.g., by atom collisions. This density restriction applies in particular to hot atomic vapors. More promising in this respect are ultracold gases~\cite{coldEIT1,coldEIT2,coldEIT3,coldEIT4,coldEIT5,coldEIT6,coldEIT7,coldEIT8,coldEIT9,coldEIT10,coldEIT11,coldEIT12,coldEIT13,coldEIT14, dense1,dense2,dense3}, in which atomic collisions are much less frequent, leading to greatly improved coherence properties. However, most experiments still operate in the regime of a dilute gas, where cooperative effects do not play a role.
But with recent advances in preparation techniques, now trapped ultracold gases at densities up to $10^{14} - 10^{15}$ atoms/cm$^3$ with a linear extend of the gas in the $\mu$m range~\cite{dense1,dense2,dense3} have been reported. With such setups, dense gas light propagation experiments at the onset of cooperative effects seem within reach.

From the theoretical side, it is well-known that in the high density regime ($N \lambda^3 \gg 1$), a rigorous treatment of cooperative effects in general gives rise to an infinite hierarchy of polarization correlations~\cite{ddQFT1,ddQFT2,ddQFT3}. Calculations of this type are demanding, such that concrete results for more complex model systems such as EIT have not been found yet. However, at the onset of cooperative effects, the infinite hierarchy can be truncated. A study of optical properties of a dense cold two-level gas found that in leading order of the density, three corrections with respect to a dilute gas occur~\cite{dd2ndorder}. 
First, the microscopic field $E_{mic}$ driving the individual atom is no longer the externally applied field $E_{ext}$ alone, but rather is corrected by the mean polarization $P$ of the neighboring atoms. This local field correction (LFC) is described by the well-known Lorentz-Lorenz formula~\cite{bornwolf}, 
\begin{align}
\label{lle}
E_{mic} = E_{ext} +  \frac{1}{3 \epsilon_0} P\,.
\end{align}
The second correction in leading order of the density is due to the quantum statistics of atoms, and becomes relevant close to the phase transition to a Bose-Einstein condensate (BEC). In this regime the atomic de Broglie wavelength is of the order of the electronic transition wavelength such that the medium's refractive and absorptive properties are enhanced by quantum mechanical exchange effects. In contrast, sufficiently above or below the transition point, quantum statistical corrections are small~\cite{dd2ndorder}. 
The third correction arises due to the leading order of multiple scattering, which can be interpreted as a reabsorption of spontaneously emitted photons before they leave the sample. In general, the reabsorption gives rise to a complicated dynamics because of the many pathways a photon can take between the different atoms in the medium. But to a first approximation, this radiation trapping can be modeled in a mean field like approach by suitably chosen additional incoherent pump rates between the ground states and the excited state~\cite{radtrapF1,radtrapF2,radtrap1,radtrap2}.

These results suggest that at the onset of cooperative effects, and at parameters away from the phase transition to BEC, the dominant cooperative corrections arise on a mean-field level, such that a macroscopic treatment is meaningful. Similar results were obtained in a recent calculation which for a two-level system explicitly compares a microscopic treatment based on a multiple-scattering approach including $n$-particle correlations with a macroscopic treatment based on LFC~\cite{sokolov}. It was found that with densities of few particles per  wavelength cubed ($N\lambda^3\approx 5$), the two methods agree well, while at higher density ($N\lambda^3\approx 124$) deviations occur, which however due to the numerical complexity could only be studied for rather small sample sizes.
So far, LFC alone has been studied in a number of systems~\cite{gmbe}, including self-induced transparency for short and ultrashort pulses~\cite{lfcsoliton,few-cycle}, coherent population trapping~\cite{agarwal-lfc} and lasing without inversion~\cite{lfclwi}.  More recently, LFC was studied in the context of atomic gases with a negative refractive index (NRI)~\cite{negindex1,negindex2,negindex3,negindex4,negindex5}. LFC effects have also been exploited to spectroscopically resolve the hyperfine structure of the Rb $D_2$ line in a dense gas~\cite{lfcspec}. 

But surprisingly, cooperative effects in EIT have received only  little attention. In~\cite{radtrapF1,radtrapF2}, effects of radiation trapping on EIT were studied. The propagation of two non-adiabatic propagating pulses is considered in~\cite{lfceit}. Some effects such as a modification of the group velocity and a phase modulation are reported, but only numerical results are given without clear interpretation~\cite{note}. 

The recent theoretical proposals for NRI requiring high density already motivate a better study of the regime of higher densities in atomic gases. But also several aspects of EIT itself could benefit strongly from such an analysis.
For example, the group velocity reduction in EIT and the related spatial compression of the pulse are directly proportional to the density~\cite{eit2}. Also, the time delay bandwidth product, a figure of merit for the overall performance of a slow light system, depends on density~\cite{dbp}. Finally, a higher density could facilitate a more efficient coupling of light and matter, e.g., for applications in quantum information science~\cite{quantinf}.

\begin{figure}[t!]
\includegraphics[width=6cm]{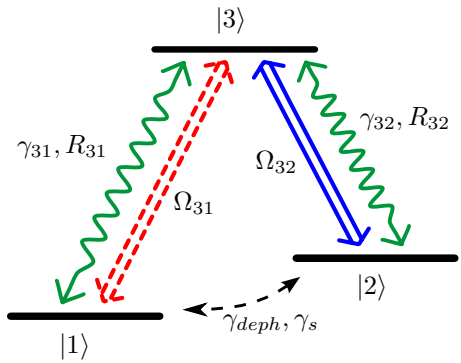}
\caption{\label{fig1}(Color online) EIT level scheme with probe field $\Omega_{31}$ (red dashed arrow) and control field $\Omega_{32}$ (blue solid arrow). The green wiggly arrows indicate the radiative decay $\gamma_{31}$, $\gamma_{32}$ and the incoherent pump rates $R_{31}$, $R_{32}$, while $\gamma_{deph}$, $\gamma_s$ denote the ground state dephasing and the ground state population transfer.}
\end{figure}

Therefore, in this paper, we study light propagation in an EIT medium at the onset of cooperative effects, and reveal and interpret the underlying physical mechanisms. We show that LFC leads to a phase modulation of a probe pulse at densities achieved in current experiments. This phase modulation is distinctively different from nonlinear self-phase modulation, as it leads to a linear frequency chirp across the whole probe pulse, and does not require high probe intensity.
Our main aim is a physical understanding of the cooperative effects. For this, we include LFC, but assume parameters away from the crossover point to a Bose-Einstein condensate, such that corrections due to quantum statistics can be neglected. Similarly, our semi-classical treatment of the light fields does not allow to reveal non-classical properties of the scattered light~\cite{quantum}. We model the onset of multiple scattering via an incoherent pump rate proportional to the excited state population. Furthermore, we take into account a ground state decoherence due to elastic collisions as well as a ground state population transfer due to spin-exchange collisions. In addition to a full numerical solution of the equations of motion (EOM), we derive analytic solutions for the propagation dynamics of a slow light pulse and the medium polarization, which  enable us to provide an intuitive explanation of the phase modulation in terms of the energy exchange with neighboring atoms.

The paper is organized as follows. In the next section (Sec.~\ref{secmod}) the theoretical model is described. We present the full set of equations of motion (EOM) (Sec.~\ref{seceom}), discuss our framework for taking into account radiation trapping (Sec.~\ref{secrad}), and introduce LFC into the EOM (Sec.~\ref{seclfc}). In Sec.~\ref{secres}, we present our results. First, we give an analytic expression for the susceptibility (Sec.~\ref{secsus}). Then, we use this expression to derive an analytic solution for a propagating probe pulse including the LFC-induced phase modulation (Sec.~\ref{secana}). In Sec.~\ref{secnum}, we compare this analytic solution to the numeric solution of the full set of equations of motion and calculate the expected probe field absorption due to radiation trapping and ground state population transfer for a range of different possible experimental parameters. In Sec.~\ref{secpha}, we explain the physical origin of the phase modulation in terms of the energy exchange of neighboring atomic dipoles and contrast it with the self-phase modulation a strong pulse acquires in a nonlinear medium. Finally, in Sec.~\ref{seccon}, we draw some conclusions.

\section{\label{secmod}Theoretical model}
\subsection{\label{seceom}Maxwell-Schr\"odinger equations}
We start from the Hamiltonian for an EIT level scheme~\cite{eit2} as shown in Fig.~\ref{fig1}. In a suitable interaction picture it is given by
\begin{align}
 H = & -\hbar (\Delta A_{22} + \Delta_{31} A_{33}) \\
\nonumber & - \frac{\hbar}{2} \left(\Omega_{31} A_{31} + \Omega_{32} A_{32} + \textrm{h.c.}\right)\,,
\end{align}
where $\Delta = \Delta_{31} - \Delta_{32}$ is the two-photon detuning, $A_{ij} = \ket{i}\bra{j}$ are the atomic operators, and $\Omega_{31}$,  $\Omega_{32}$, $\Delta_{31}$, $\Delta_{32}$ are the Rabi frequency and detuning of the probe and control field, respectively. Note that without application of the LFC, the Rabi frequencies contain microscopic fields at the position of the respective atoms. After performing LFC, in our numerical calculations, the microscopic electric fields are replaced by the externally applied fields as discussed in Sec.~\ref{seclfc}.

In order to account for the various incoherent processes, we choose a  master equation approach~\cite{FiSw2005}, and the EOM for the atomic density operator $\rho$ reads
\begin{subequations}
\begin{align}
\label{eqn:mas}
\partial_t \rho = & \, \frac{1}{i \hbar} \left[ H, \rho \right] \\
\nonumber & - \sum_{j=1}^2 \frac{\gamma_{3j}}{2} \left( \com{\rho A_{3j}}{A_{j3}} + \textrm{h.c.} \right) \\
\nonumber & - \sum_{j=1}^2 \frac{R_{3j}}{2} \left( \com{A_{3j}}{\com{A_{j3}}{\rho}} + \textrm{h.c.} \right) \\
\nonumber & - \frac{\gamma_s}{2} \left( \com{A_{21}}{\com{A_{12}}{\rho}} + \textrm{h.c.} \right) \\
\nonumber & - \gamma_{deph} \left(A_{22} \rho A_{11} + \textrm{h.c.} \right)\,.
\end{align}
Here, $\gamma_{3j}$ ($j \in \{1,2\}$)  denote the radiative decay rates, and and $R_{3j}$ are incoherent pumping rates between the ground and the excited state. $\gamma_s$ is the ground state population transfer rate, and $\gamma_{deph}$ is the ground state dephasing.

The dynamics of the probe and control fields are each governed by a wave equation. In slowly varying envelope approximation (SVEA), these have the form~\cite{scullybook}
\begin{align}
\label{eqn:wav}
\left[\partial_z + \frac{1}{c} \partial_t \right] \mathcal{E}_{3j} =  \frac{i k}{\varepsilon_0}\,P_{3j} \,, \quad j \in \{1,2\}\,,
\end{align}
\end{subequations}
where $c$ is the speed of light in vacuum, $\mathcal{E}_{3j}$ is the envelope of the probe and control field, and $P_{3j}$ is the probe and control field polarization.

The combination of Eq.~(\ref{eqn:mas}) and Eqs.~(\ref{eqn:wav}) form the full set of Maxwell-Schr\"odinger EOM that govern the spatio-temporal dynamics of the system. In our model, its solution becomes more challenging because of the additional nonlinear terms introduced due to radiation trapping and LFC. We will discuss both aspects separately and in more detail in the following.

\subsection{\label{secrad}Radiation trapping}
A higher density of the atoms leads to reabsorption and multiple scattering of spontaneously emitted photons. We model this by incoherent pump fields $R_{31}$ and $R_{32}$ with an intensity proportional to the population of the excited state. The number of incoherent photons inside the sample is also influenced by its geometry. Along the lines of~\cite{radtrap1}, this is taken into account by a pumping rate $r_a$ due to atomic decay and a photon escape rate $r_e$. Altogether, $R_{31}$ and $R_{32}$ are then given by
\begin{subequations}
\begin{align}
 R_{31} = \gamma_{31} \frac{r_a}{r_e} \rho_{33}\,, \\
 R_{32} = \gamma_{32} \frac{r_a}{r_e} \rho_{33}\,.
\end{align}
\end{subequations}
Obviously, the ratio $r_a/r_e$ should also depends on the atomic density. For high density it will eventually approach unity such that even a small fraction of excited state population can lead to a significant change of the dynamics. In an EIT system, the excited state population in principle should remain small. Especially under adiabatic conditions when the system is in the dark state at all times, the destructive quantum interference underlying EIT strongly suppresses excitation events. However, in experiments a dephasing of the ground state coherence as well as spin exchange due to inhomogeneous magnetic fields and atomic collisions can lead to an effective pump rate into the bright state and subsequent excitation. This type of dynamics is modeled by the different incoherent pump rates in the master equation Eq.~(\ref{eqn:mas}). In Sec.~\ref{secnum}, we use the numerical solution of the full set of EOM including this dynamics to calculate the probe field absorption for a range of different values for $r_a/r_e$ as well as $\gamma_s$.

\subsection{\label{seclfc}Local field corrections}
In the master equation (\ref{eqn:mas}) the Rabi frequencies depend on the microscopic probe and control field, such that we have to replace them by their macroscopic counterparts using Eq.~(\ref{lle}). Since the mean medium polarization $P$ can be expressed in terms of a density matrix element, this leads to nonlinear EOM. Expanding the nonlinear EOM up to linear order in the external probe field leaves us with just two EOM, one for the probe field coherence $\rho_{31}$ and one for the Raman coherence $\rho_{21}$,
\begin{subequations}
\begin{align}
\label{eom:1}
\partial_t \rho_{31} = & - \Gamma_{31} \rho_{31} + \frac{i}{2} \Omega_{31} + \frac{i}{2} L \gamma_{31} \rho_{31} + \frac{i}{2} \Omega_{32} \rho_{21}\,,\\
\label{eom:2}
\partial_t \rho_{21} = & - \Gamma_{21} \rho_{21} + \frac{i}{2} \Omega_{32}^* \rho_{31}\,.
\end{align}
\label{eom}
\end{subequations}
From now on, after performing the LFC, $\Omega_{31}$ and $\Omega_{32}$ denote the respective Rabi frequencies of the external fields, and we have defined
\begin{subequations}
\begin{align}
\Gamma_{31} = & \frac{\gamma}{2} - i \Delta_{31}, \\
\Gamma_{21} = & \gamma_{dec} - i \Delta, \\
\gamma = & \gamma_{31} + \gamma_{32} + \gamma_s, \\
\gamma_{dec} = & \gamma_{deph} + \gamma_s\,.
\end{align}
\end{subequations}
Due to LFC, a new term arises in Eq.~(\ref{eom:1}) as compared to the low-density case which is proportional to the dimensionless parameter
\begin{align}
 L = \frac{N \lambda^3}{4 \pi^2}.
\end{align}
It is a measure for the strength of LFC, and the prefactors ensure that $L = 1$ corresponds to a density where the LFC induced frequency shift in a two-level atom is equal to half the natural linewidth. Formally, the new term can be interpreted as a frequency shift in an EIT system as well and can be included into the probe field detuning $\tilde{\Delta}_{31} = \Delta_{31} + L \gamma_{31}/2$. However, this frequency shift does not influence the two-photon detuning $\Delta$.

\section{\label{secres}Results}
\subsection{\label{secsus}Susceptibility with LFC}
\begin{figure}[t!]
\includegraphics[width=\columnwidth]{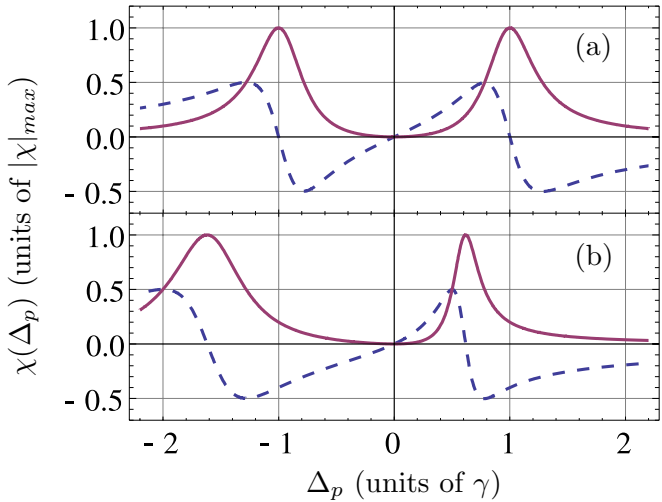}
\caption{\label{fig2}(Color online) Real part (blue dashed line) and imaginary part (red solid line) of the EIT susceptibility given by Eq.~\ref{sus} for (a) negligible LFC ($L = 10^{-5}$) and (b) strong LFC ($L = 4$).}
\end{figure}

The EOM~(\ref{eom}) are already linear in the probe field and solving for the steady state, we can derive the susceptibility $\chi$,
\begin{align}
\label{sus}
\chi = \frac{3 L \frac{\gamma_{31}}{2} (\Delta + i \gamma_{dec})}{\frac{\gamma}{2} \gamma_{dec} - \tilde{\Delta}_{31} \Delta + \frac{|\Omega_{32}|^2}{4} - i (\tilde{\Delta}_{31} \gamma_{dec} + \Delta \frac{\gamma}{2})}\,.
\end{align}
With an analytic expression at hand we can easily pinpoint the effect of LFC. We find that a reshaping of the EIT transparency window takes place~\cite{agarwal-lfc}. To illustrate this, in Fig.~\ref{fig2} we show the real and imaginary part of $\chi$ for two different sets of parameters. In Fig.~\ref{fig2}(a) we set $L = 10^{-5}$ which results in negligible LFC and the standard form of the EIT susceptibility is recovered. With $L = 4$ in Fig.~\ref{fig2}(b) a strong reshaping of the transparency window due to LFC is found. In both cases the control field Rabi frequency is $\Omega_{32} = 2 \gamma$ and the control field detuning is $\Delta_{32} = 0$.

%

\subsection{\label{secana}Analytic solution for a light pulse}
To analyze the influence of LFC on the propagation dynamics of a light pulse, we expand Eq.~(\ref{sus}) around the center of the transparency window. From 
\begin{align}
 k(\omega) = \frac{\omega}{c} \sqrt{1+\chi(\omega)}\,,
\end{align}
we find the frequency dependent wave number $k(\omega)$ up to second order in the probe field detuning. The solution for the positively rotating component of the probe field in Fourier space then follows from
\begin{align}
 \label{solfou}
E^{(+)}(z, \omega) = E^{(+)}(0, \omega) \exp[i k(\omega) z],
\end{align}
where $E^{(+)}(0, \omega)$ is given by the initial condition and 
\begin{align}
 \label{wavenumber}
 k(\omega) \approx k_0 + i \frac{ n_g \gamma_{dec}}{c} + \frac{\Delta_{31}}{v_g}  + k_0 (i \beta_1+ \beta_2 ) \Delta_{31}^2 \,.
\end{align}
We neglected terms suppressed by a factor of $\gamma \gamma_{dec}/\Omega_{32}^2$, since $\gamma \gamma_{dec} \ll \Omega_{32}^2$ is required for low absorption, and of higher order in $\Delta_{31}$. Each term in Eq.~(\ref{wavenumber}) can be clearly interpreted.
$k_0$ is the wave number of the undisturbed carrier wave. The second term describes the decay due to the ground state decoherence $\gamma_{dec}$ where 
\begin{align}
n_g = \frac{3 L \gamma_{31} \omega_0}{\Omega_{32}^2}
\end{align}
is the group index and $\omega_0$ is the probe field transition frequency. The third term leads to the reduced group velocity $v_g = c / (1 + n_g)$. The fourth term is quadratic in $\Delta_{31}$ and thus associated with a change of width in the Fourier transformation of a Gaussian. The imaginary part proportional to 
\begin{align}
\beta_1 = \frac{6 L \gamma \gamma_{31}}{\Omega_{32}^4}
\end{align}
leads to a broadening of the temporal width due to the finite spectral width of the transparency window. Finally, the real part proportional to
\begin{align}
\beta_2 = \frac{6 L^2 \gamma_{31}^2}{\Omega_{32}^4}
\end{align}
results from LFC. Formally, it leads to an imaginary part in the temporal width which corresponds to a phase modulation of the pulse. When we assume a Gaussian pulse shape for the initial condition, the solution in the time domain can be obtained by Fourier transformation. Considering only the envelope defined by
\begin{align}
 E(z,t) = \frac{1}{2} \mathcal{E}(z,t) \exp[i(k_0 z - \omega_0 t)] + \textrm{c.c.}\,,
\end{align}
we find
\begin{align}
\label{soltim}
\mathcal{E}(z,t) = \mathcal{E}_0 \frac{\sigma}{\tilde{\sigma}} \exp\left[-\gamma_{dec} \frac{z}{v_g } - \frac{(t-z/v_g)^2}{2\tilde{\sigma}^2}\right]\,.
\end{align}
$\mathcal{E}_0$ and $\sigma$ are the initial amplitude and temporal width. The LFC modified width  after propagating a distance $z$ is
\begin{align}
 \tilde{\sigma}^2 = \sigma^2 + 2 k_0 z (\beta_1 - i \beta_2).
\end{align}
The phase modulation can be well approximated by the parabola
\begin{align}
 \label{phlfc}
 \phi_\textrm{\tiny{LFC}}(t) = \frac{\beta_2 k_0 z}{\sigma^2} \left[1 - \left(t-z/v_g\right)^2/\sigma^2\right].
\end{align}

\subsection{\label{secnum}Numeric solution}
To compare our analytic solution to results obtained numerically, we chose parameters consistent with recent ultra cold gas setups~\cite{dense1,dense2,dense3}. We assume a medium with density $N \approx 10^{14} \textrm{cm}^{-3}$ and a length of $z \approx 40 \mu$m. This corresponds to $N \lambda^3 \approx  50$ ($\lambda = 795$nm), and with a control field $\Omega_{32} = 2 \gamma$, a probe pulse with initial width of $\sigma = 20 / \gamma$ propagates a distance of $z = 150 v_g / \gamma$. Besides some broadening due to the finite spectral width of the transparency window, the probe pulse is attenuated directly by the ground state decoherence $\gamma_{dec}$ (see Eq.~\ref{soltim}). It also suffers absorption because of the reabsorption of incoherent photons. As discussed in Sec.~\ref{secrad}, this significantly depends on the ground state population transfer $\gamma_s$ and the geometry of the sample represented by $r_a/r_e$.

\begin{figure}[t!]
\includegraphics[width=\columnwidth]{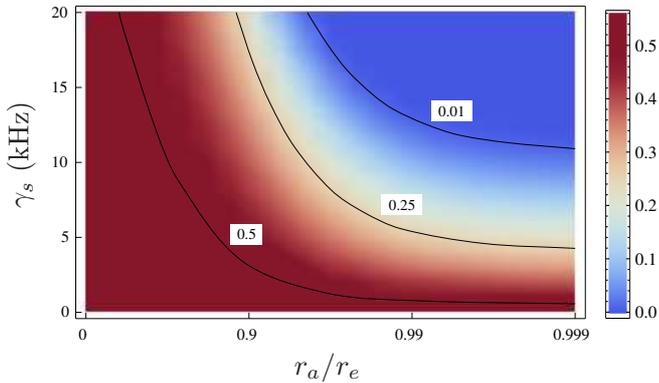}
\caption{\label{fig3}(Color online) Probe field absorption in dependence of the radiation trapping rate $r_a/re$ and the ground state transfer rate $\gamma_s$. The shading indicates the peak value of the propagated probe field amplitude $\Omega_{31}(z)$ relative to the initial amplitude $\Omega_{31}(0)$. The three contour lines indicate parameters at which $50\%$, $25\%$ or $1\%$ of the initial amplitude leave the sample. The results are obtained using the numerical solution of the full set of EOM. All other parameters remain fixed and are chosen as discussed in Sec.~\ref{secnum}.}
\end{figure}

From results of a recent experiment with Rubidium atoms of density $N\approx 2\times10^{14} \textrm{cm}^{-3}$ in an anisotropic trap we estimate $\gamma_{dec}$ to about $100$~kHz~\cite{dense1,dense2,dense3}. From radiation trapping experiments in hot and ultracold gases~\cite{radtrap1,radtrap2}, we further estimate that the ratio of the photon pump and escape rate is about $r_a/r_e = 0.99$. The rate of spin exchange collisions depends on the relative ground state energies and the fact that spin exchange collision become more likely if the overall angular momentum is conserved. From \cite{coldfermions} we find that a typical value in ultracold gases is given by
\begin{align}
 \gamma_s = N \times 10^{-11} \textrm{cm}^3 \textrm{s}^{-1}\,,
\end{align}
such that for a density of $N \approx 10^{14} \textrm{cm}^{-3}$ we estimate at a value of about $\gamma_s = 1$~kHz. To account for different experimental situations, and to obtain insight in the effect of radiation trapping and ground state decoherence, we use the full set of EOM to numerically propagate a probe pulse and to calculate the probe field absorption for a range of values for $\gamma_s$ as well as $r_a/r_e$. The result is shown in Fig.~\ref{fig3}. The plot shows the amplitude of the propagated pulse relative to the initial pulse. The three contour lines indicate parameters for which a fraction of $50\%$, $25\%$ or $1\%$ of the initial light field amplitude leaves the sample, respectively. For the estimated values of $\gamma_{dec}$, $\gamma_s$, and $r_a/r_e$ which we assume for our further calculations, the probe field $\Omega_{31}$ is reduced to about $50\%$ of its initial value.

\begin{figure}[t!]
\includegraphics[width=\columnwidth]{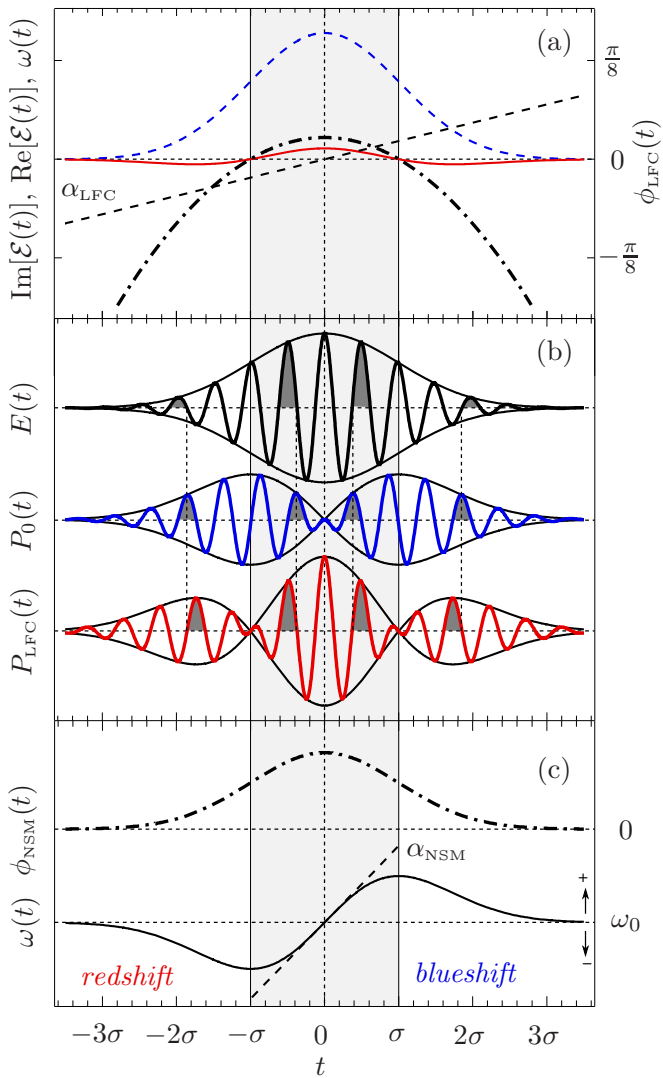}
\caption{\label{fig4}(Color online) 
(a) Real part (blue dashed line) and imaginary part (red solid line) of a Gaussian pulse with LFC induced phase modulation. Both parts are scaled to arbitrary but equal units. The parabola (black dash-dotted line) shows the time dependent phase shift in radian measure (right axis) and the dashed straight line is the corresponding instantaneous frequency. It exhibits a linear frequency chirp with slope $\alpha_\textrm{\tiny{LFC}}$ over the total extend of the pulse. 
(b) Envelope and carrier wave of the pulse (black upper line), the polarization induced directly by the pulse (blue middle line), and the additional polarization induced by dipole-dipole interactions (red lower line). The amplitudes are not drawn to scale and the carrier wave length has been strongly exaggerated to make phase relations clearly visible. 
(c) The nonlinear self-phase modulation (black dash-dotted line) of a Gaussian pulse. The frequency chirp is approximately linear only in the pulse center (black dashed straight line) with slope $\alpha\textrm{\tiny{NSM}}$.}
\end{figure}

Fig.~\ref{fig4} analyzes the pulse after passing through the medium. In subfigure~\ref{fig4}(a), we show the propagated pulse together with the LFC induced phase modulation $\phi_\textrm{\tiny{LFC}}(t)$ and the corresponding instantaneous frequency defined by $\omega(t) = \omega_0 - \partial_t \phi_\textrm{\tiny{LFC}}(t)$. The numerical solution is virtually indistinguishable from the analytical one, and we only show the analytical solution. We see that the phase modulation has a negative parabola-like shape such that the instantaneous frequency is approximately linear over the total extend of the pulse.

%

\subsection{\label{secpha}Physical origin of the phase modulation}
To explain the physical origin of the phase modulation, we explicitly calculate the relevant parts of the polarization using the relation 
\begin{align}
 P^{(+)}(z, \omega) = \epsilon_0 \chi(\omega) E^{(+)}(z, \omega)\,.
\end{align}
Considering only the real part of $\chi$ up to quadratic order in $\Delta_{31}$ and Fourier transforming it into the time domain, we can distinguish two contributions,
\begin{subequations}
\begin{align}
 &P^{(+)}_0(z, t) = \frac{\epsilon_0 n_g}{\omega_0} \left[i \partial_t \mathcal{E}(z, t)\right] \exp[i(k_0 z - \omega_0 t)] \label{contr:1}\,,\\
 &P^{(+)}_\textrm{\tiny{LFC}}(z, t) = \epsilon_0 \beta_2 \left[i^2 \partial^2_t \mathcal{E}(z, t)\right] \exp[i(k_0 z - \omega_0 t)]\,. \label{contr:2}
\end{align}
\label{contr}
\end{subequations}
The first contribution stems from the part linear in $\Delta_{31}$ and leads to the change of group velocity. The second contribution is due to the part quadratic in $\Delta_{31}$ which is related to the LFC induced phase modulation. In Fig.~\ref{fig4}(b), we show the pulse together with these two contributions.
In the first half of the pulse, $E(t)$ is ahead in phase by $\pi/2$ compared to $P_0(t)$, which indicates that energy is transferred from the pulse to the polarization $P_0(t)$. In the second half, the pulse is delayed by $\pi/2$, and energy is transferred back from the polarization $P_0(t)$ to the pulse. This energy exchange effectively reduces the group velocity of the pulse.
Similarly, we can understand how the interaction of atoms with the collective dipole field of their neighbors proportional to $P_0(t)$ induces an additional polarization $P_\textrm{\tiny{LFC}}(t)$. Before $t = -\sigma$, the polarization component $P_0(t)$ is $\pi/2$ ahead in phase compared to $P_\textrm{\tiny{LFC}}(t)$, whereas at $-\sigma < t < 0$, it is delayed by $\pi/2$, again leading to an energy exchange. The same exchange takes place again for $0 < t < \sigma$ and $t > \sigma$.
While the additional polarization $P_\textrm{\tiny{LFC}}(t)$ is induced exactly by the same mechanism as $P_0(t)$, its back action on the probe pulse is different. At $t < -\sigma$ and $t > \sigma$,  $P_\textrm{\tiny{LFC}}(t)$ and $E(t)$ have opposite phase. This opposite phase has a dragging effect on $E(t)$, and reduces its phase. In the central part of the pulse ($-\sigma < t < \sigma$), $P_\textrm{\tiny{LFC}}(t)$ is in phase with $E(t)$. This has a pushing effect on $E(t)$, and increases the phase of the pulse. This interpretation agrees with the phase modulation obtained from the calculation as shown in Fig.~\ref{fig4}(a). 
We thus conclude that energy is exchanged between the atomic dipoles and the field of neighboring dipoles in exactly the same way as between the atomic dipoles and the external field $E(t)$. But the two polarization components act differently on the probe pulse, leading to the group velocity change and the phase modulation, respectively.

Finally, we compare the LFC induced phase modulation with nonlinear self-phase modulation (NSM) in a medium with an intensity dependent refraction. The NSM modulation is 
\begin{align}
 \phi_\textrm{\tiny{NSM}}(t) = n_2\, I(t)\, k_0\, z\,,
\end{align}
where $n_2$ is the intensity dependent index of refraction and $I(t)$ the intensity profile of the pulse. In Fig.~\ref{fig4}(c) we show $\phi_\textrm{\tiny{NSM}}(t)$ together with the corresponding instantaneous frequency for a Gaussian pulse. We see that the front of the pulse experiences a red shift whereas the back experiences a blue shift with an approximately linear frequency chirp in the center. Comparing the two chirps,
\begin{align}
 \alpha_\textrm{\tiny{NSM}} = & \frac{2 n_2 I_0 k_0 z}{\sigma^2}\,, \\
 \alpha_\textrm{\tiny{LFC}} = & \frac{2 \beta_2 k_0 z}{\sigma^4}\,,
\end{align}
we find that in the LFC case, $n_2 I_0$ is replaced by $\beta_2/\sigma^2$. Thus, the LFC modulation does not require a large intensity, and is approximately linear over the total extend of the pulse since it depends on the strength of the dipole-dipole interaction. This strength is given by $\beta_2$ in an EIT system and can be influenced by the density and the control field strength $\Omega_{32}$.

\section{\label{seccon}Conclusion}
In summary, we studied light propagation in an EIT system for densities at the onset of cooperative effects.
For this, we amended the standard Maxwell-Schrödinger equations for a three-level $\Lambda$-type EIT setup on a mean-field level by local field corrections of the applied fields, and by a model for radiation trapping based on incoherent pump rates between the ground and the excited states.
As main result, we found that the phase of a Gaussian probe pulse is strongly modulated by the polarization contribution to the local field.
We gave a physical interpretation for the underlying mechanism leading to the phase modulation and showed that it is distinctively different from the case of non-linear self phase modulation, as the resulting frequency chirp is linear across the whole pulse, and does not depend on the pulse intensity.
Based on the numerical solution, we further studied the effects of decoherence and radiation trapping. Within our model, this allows to estimate upper limits for these incoherent effects such that the pulse survives a sufficient propagation length for the phase modulation to become observable.



\begin{thebibliography}{99}
\bibitem{note}
Our results, however, do not show the dependence of the group velocity.
\end{thebibliography}
\end{document}